\documentclass[]{aa}  

\usepackage{natbib}
\bibpunct{(}{)}{;}{a}{}{,}
\usepackage{graphicx}
\usepackage{revsymb}	
\usepackage{amsmath}	
\usepackage[usenames]{color}	
\usepackage{epstopdf}	
\usepackage{txfonts}
\usepackage{amssymb}
\usepackage{color}
\usepackage[justification=centering]{caption}
\usepackage{geometry} 
\usepackage[parfill]{parskip} 
\usepackage{amssymb}
\usepackage{slantsc}
\usepackage{array}
\usepackage{url}
\usepackage{amsfonts}
\usepackage[colorlinks=true,linkcolor=black, urlcolor=black, citecolor=black]{hyperref}
\usepackage{hyperref} 
\usepackage{sidecap}
\usepackage{graphicx}
\usepackage{xcolor}
\usepackage{nicefrac}
\usepackage{subfigure}
\usepackage{epsfig}
\colorlet{rouge}{red!70!darkgray}

\begin{document}
\title{Constraining convective regions with asteroseismic linear structural inversions}
\author{G. Buldgen\inst{1,2}\and D. R. Reese\inst{3}\and M. A. Dupret\inst{1}}
\institute{Institut d’Astrophysique et Géophysique de l’Université de Liège, Allée du 6 août 17, 4000 Liège, Belgium \and School of Physics and Astronomy, University of Birmingham, Edgbaston, Birmingham B15 2TT, United Kingdom \and LESIA, Observatoire de Paris, PSL Research University, CNRS, Sorbonne Universités, UPMC Univ. Paris 06, Univ. Paris Diderot, Sorbonne Paris Cité, 5 place Jules Janssen, 92195 Meudon Cedex, France}
\date{February 2017}
\abstract{Convective regions in stellar models are always associated with uncertainties, for example due to extra-mixing or the possible inaccurate position of the transition from convective to radiative transport of energy. Such inaccuracies have a strong impact on stellar models and the fundamental parameters we derive from them. The most promising method to reduce these uncertainties is to use asteroseismology to derive appropriate diagnostics probing the structural characteristics of these regions.}
{We wish to use custom-made integrated quantities to improve the capabilities of seismology to probe convective regions in stellar interiors. By doing so, we hope to increase the number of indicators obtained with structural seismic inversions to provide additional constraints on stellar models and the fundamental parameters we determine from theoretical modeling.}
{First, we present new kernels associated with a proxy of the entropy in stellar interiors. We then show how these kernels can be used to build custom-made integrated quantities probing convective regions inside stellar models. We present two indicators suited to probe convective cores and envelopes, respectively, and test them on artificial data.}
{We show that it is possible to probe both convective cores and envelopes using appropriate indicators obtained with structural inversion techniques. These indicators provide direct constraints on a proxy of the entropy of the stellar plasma, sensitive to the characteristics of convective regions. These constraints can then be used to improve the modeling of solar-like stars by providing an additional degree of selection of models obtained from classical forward modeling approaches. We also show that in order to obtain very accurate indicators, we need $\ell=3$ modes for the envelope but that the core-conditions indicator is more flexible in terms of the seismic data required for its use.}
{}
\maketitle
\section{Introduction} \label{SecIntro}
Inversion techniques have been used for several decades in helioseismology to analyze the structure of the Sun. Amongst the greatest successes of this field, one finds the determination of the base of the solar convective envelope \citep{KosConv} and the helium abundance in this region \citep{AntiaY, KosovY}, as well as the inversion of the sound speed profile \citep{AntiaC}. While for the Sun it is possible to determine a whole structural profile with linear techniques, the case of asteroseismic inversions is far more difficult. This is due to the small number of modes and the absence of oscillations of high harmonic degree, which can help to scan through the whole structure of the star. Initial studies have been performed to carry out inversions based on the variational principle for stars other than the Sun \citep[See][for a few examples of such studies.]{GoughAster, GoughPmod, Takata, BasuAster}.

Since the advent of space photometry missions such as CoRoT \citep{Baglin} and Kepler \citep{Borucki}, we now have seismic data with sufficiently small uncertainties to attempt structural inversions to constrain models of solar-like stars. In the future, the TESS and Plato $2.0$ missions \citep{Rauer} will bring further data, continuing what is now named the space photometry revolution.

While for some of the best Kepler targets, full profile inversions could be attempted, they would require very specific conditions. For example, a convective parameter inversion as shown in \citet{KosovY} would need a very accurate knowledge of the mass and radius of the target as shown in \cite{BuldgenKer} from the analysis of the linear relations for various structural kernels. Secondly, an efficient way to deal with surface corrections without degrading the capabilities of the inversion technique would need to be found. As a workaround to these difficulties, and to allow more versatile applications of these techniques, \citet{Reese} adapted the classical inversion techniques so that they would focus on extracting global information such as the mean density from the oscillation spectra. This global information is defined by integrated quantities which are chosen for their particular ability to probe certain aspects of stellar structure \citep[see for example]{Buldgentau,Buldgentu}.

The strength of this approach is that it focuses all of the information provided by the seismic spectra on the determination of one piece of information at a time. Moreover, this information, which is a linear combination of frequency differences is related through the variational integral relations \citep{Dziembowski} to structural characteristics. Our goal is to further correct seismic models which have been built using the classical forward modeling method used in asteroseismology. Once the integrated quantity is defined, one can use the SOLA inversion technique \citep{Pijpers} to check whether it is possible to obtain corrections of the chosen structural indicator. The success of the operation depends on whether the integrated quantity behaves linearly and whether the target function can be easily fitted with the amount of seismic information available. Ultimately, one still faces the classical trade-off problem of inversion techniques and the fine-tuning of the parameters has to be done carefully if one wishes to efficiently extract the structural information from the seismic observations. 

In this paper, we present results from tests on artificial data for new indicators based on an entropy proxy. We start by presenting the kernels associated with the structural quantity we use in the inversions. Furthermore, we show how this variable naturally reproduces the entropy plateaus in adiabatic convective regions and how its behavior can be used to probe both convective cores and envelopes with custom-built indicators. Probing convective regions and their surrounding layers is crucial as they are likely subject to extra-mixing.

These questions are illustrated by the current uncertainties on the solar tachocline \citep[see][and references therein]{Zahn} and emphasize the physical complexity linked to convective envelopes. In the Sun, indications of mixing can be seen in the relative sound speed differences or from rotation inversions, where the change from differential to solid body rotation is associated with this particular region of solar structure. In asteroseismology, convective penetration has also been observed in the CoRoT target HD52265 \citep{Lebreton}. From a seismic point of view, glitch fitting techniques may help to position the acoustic depth of the convective envelope \citep[see][]{MonteiroOne,MonteiroTwo,Verma}. However, these techniques also require very high data quality and currently, the signal related to the base of the convective zone is considered by some authors to be too weak to be fully exploited \citep{Verma}. In that sense, our approach takes a different path, by focusing on global information rather than localized signatures to provide additional constraints.

Besides additional mixing processes, inaccuracies in the physical ingredients and numerical techniques used in stellar models can leave their mark on the boundaries of convective regions and therefore on the whole stratification of the model \citep{Gabriel}. This, in turn, will adversely affect age determinations, and to a lesser degree, the accuracy with which other stellar parameters are obtained. For example, the inaccuracy with which the extent of a convective core is determined induces uncertainties on the age of an observed target that far exceed $10\%$, emphasizing the importance of constraining convective regions in the current context of the Plato $2.0$ mission.

In Sect. \ref{SecNumerics}, we test the accuracy of both indicators for various targets and analyze their error contributions to see whether the inversion can be computed for the best solar-like targets at hand. Finally, we conclude by summarizing our results and comment on further analyses which have to be carried out to fully assess the potential of structural inversions in asteroseismology. 

\section{Kernels for the entropy proxy} \label{NewKernels}
Obtaining a new indicator with asteroseismic inversions means finding a new way to efficiently extract information from the frequencies. Due to the small number of observed modes, the target functions associated with the indicators must be constructed so that they can easily be fitted by structural kernels. Moreover, the choice of the structural variable must be physically motivated. For example, the use of the squared isothermal sound speed $u$ in \cite{Buldgentu} to analyze deep regions is motivated by the approximate relation $u\approx \frac{T}{\mu}$ in the core. In \cite{Dziembowski}, its use was motivated by the problem of measuring the helium abundance in the solar convective envelope.

In this study, we focus on convective regions. Specifically, we wish to be able to analyze the uncertainties linked to the detection of convective cores and extra-mixing at the boundaries of convective regions. However, we have to keep in mind the intrinsic limitations of inversions in asteroseismology. From our previous studies \citep{BuldgenKer}, we know that kernels like the $(A,\Gamma_{1})$ or $(A,Y)$ kernels,  where $A= \frac{d \ln \rho}{d \ln r}-\frac{1}{\Gamma_{1}}\frac{d \ln P}{d \ln r}$, cannot be used without a very accurate and precise determination of the radius, which far exceeds observational uncertainties which are of the order of a percent in the case of interferometry. This means that we have to find a new structural pair which efficiently probes convective regions. In this section, we derive new kernels associated with the variable $S_{5/3}=\frac{P}{\rho^{5/3}}$, with $P$ the pressure and $\rho$ the density, and justify their choice as an efficient probe of convective regions.

The choice of $S_{5/3}$ as a structural variable stems from its relation with the entropy of an ideal gas. Using thermodynamical relations \citep[See e.g.][for thermodynamical relations from which the following equation can be derive]{Kippenhahn}, one can show that
\begin{align}
S=\frac{3k_{B}}{2}\left(\mu m_{u} \ln \left( \frac{P}{\rho^{5/3}}\right)+f\left( \mu \right) \right),
\end{align}
with $k_{B}$ being the Boltzmann constant, $\mu$ the mean molecular weight and $m_{u}$ the atomic mass unit. In this equation, $f$ only depends on the mean molecular weight and various physical constants. 

This quantity has the interesting property of forming a plateau in the adiabatic convection zones. The height of this plateau is related to the temperature and mean molecular weight gradients in the vicinity of the convective zone's boundary and thus to the stratification of these poorly modeled regions. We illustrate this property in Sect. \ref{NewIndic}, when we derive the seismic indicators.

Due to the limited number of frequencies in asteroseismology, the second variable of the structural pair must be chosen so that the cross-term is naturally small. Two variables satisfy this condition: $\Gamma_{1}$, the adiabatic exponent defined as $\Gamma_{1}=\left(\frac{\partial \ln P}{\partial \ln \rho} \right)_{S}$ and $Y$, the helium abundance. %For $\Gamma_{1}$, it is due to the fact that $\frac{\delta \Gamma_{1}}{\Gamma_{1}}$ will naturally be smaller than other thermodynamic variables in deep regions. For $Y$, it is a consequence of the naturally small amplitude of the $Y$ kernels when compared to other structural kernels.
This means that the structural pairs we are aiming for are the $(S_{5/3},\Gamma_{1})$ and $(S_{5/3},Y)$ pairs, which can be derived from the $(\rho,\Gamma_{1})$ and $(\rho,Y)$ pairs. One could also use the $(u,\Gamma_{1})$ and $(u,Y)$ pairs to derive the differential equations without any further difficulties. However, since in practice the $(u,Y)$ and $(u,\Gamma_{1})$ kernels are already obtained from the numerical resolution of a second order differential equation, it is wiser to use the $(\rho,\Gamma_{1})$ and $(\rho,Y)$ kernels as a starting point to avoid multiplying the sources of numerical errors.

Using the direct method presented in \citet{BuldgenKer}, the equation obtained for the $(S_{5/3},\Gamma_{1})$ kernels is a second order differential equation written as follows
\begin{align}
-y \frac{d^{2}\mathcal{K}^{'}}{dy^{2}} & -\left[ 3 - \frac{2 \pi y^{3/2} \rho}{m} \right] \frac{d \mathcal{K}^{'}}{dy} = \frac{5y}{3}\frac{d^{2} \mathcal{K}}{dy^{2}} \nonumber \\ &-\left[5-\frac{\rho m}{2 P y^{1/2}} - \frac{10 \pi y^{3/2} \rho}{3m} \right] \frac{d \mathcal{K}}{dy} \nonumber \\ & -\left[ \frac{m}{4y^{1/2} \rho}\frac{d \rho}{dy} +\frac{\rho^{2}m^{2}}{4y^{2}P^{2}} - \frac{m \rho}{4P y^{3/2}}\right] \mathcal{K}, \label{eqKerS},
\end{align}
with $y=(\frac{r}{R})^{2}$ where $r$ is the radial position, $R$ the total radius of the star, and $m$ the mass contained within the sphere of radius $r$, $\mathcal{K}=K^{n,\ell}_{S_{5/3},\Gamma_{1}}$,  $\mathcal{K}^{'}=K^{n,\ell}_{\rho,\Gamma_{1}}$. As was the case for the $(u,Y)$ kernels, one can use exactly the same equation to obtain the $(S_{5/3},Y)$ kernels from the $(\rho,Y)$ kernels by simply taking $\mathcal{K}^{'}=K^{n,\ell}_{\rho,Y}$ and $\mathcal{K}=K^{n,\ell}_{S_{5/3},Y}$.

\begin{figure*}[t]
	\centering
		\includegraphics[width=17.5cm]{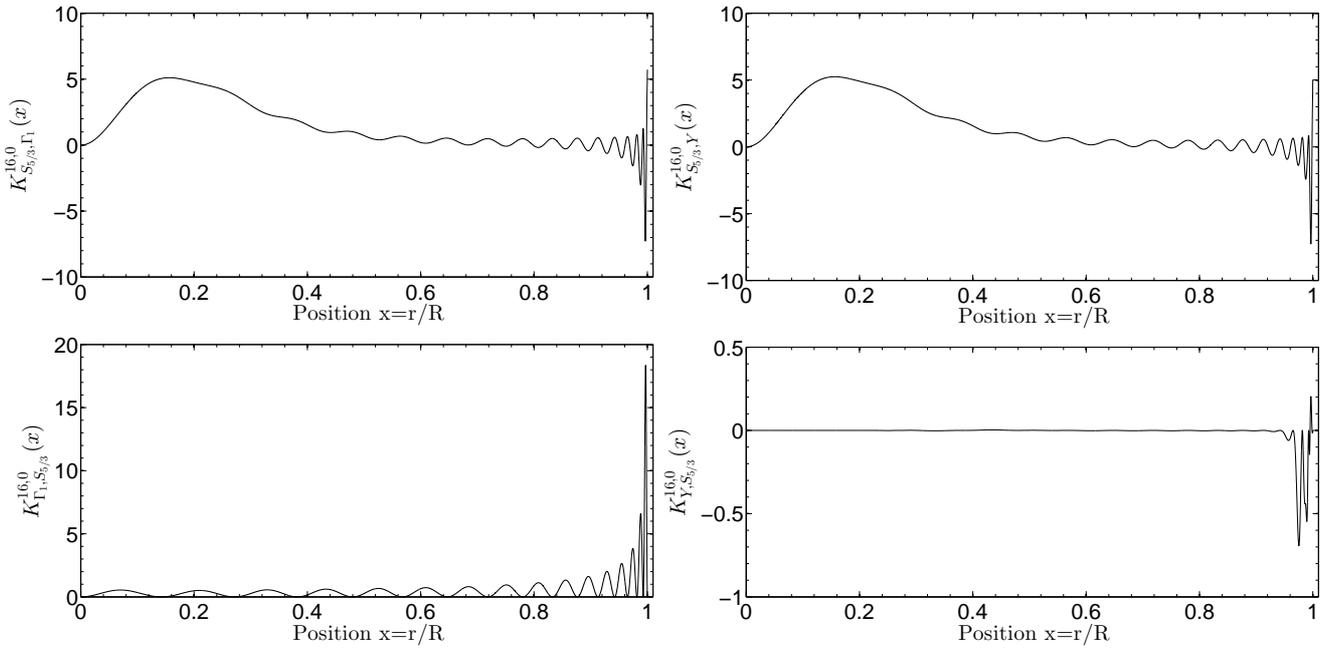}
	\caption{Structural kernels for the $n=15,\ell=0$ mode associated with the $(S_{5/3},\Gamma_{1})$ structural pair on the left-hand side and with the $(S_{5/3},Y)$ pair on the right-hand side for Target $1$ of Table \ref{tabStrucTarget}. The kernels are presented in their non-dimensional form.}
		\label{figKerS}
\end{figure*}

The $K^{n,\ell}_{\Gamma_{1},S_{5/3}}$ and $K^{n,\ell}_{Y,S_{5/3}}$ are directly obtained from the following algebraic relations:
\begin{align}
K^{n,\ell}_{\Gamma_{1},S_{5/3}} & = K^{n,\ell}_{\Gamma_{1}, \rho}, \\
K^{n,\ell}_{Y,S_{5/3}} & = K^{n,\ell}_{Y, \rho}.
\end{align}
Examples of such kernels are illustrated in Fig. \ref{figKerS}. 

The integral equation used as a starting point for the problem is used to derive the additional boundary condition required to solve the second order equation. This equation is written
\begin{align}
K^{n,\ell}_{\rho,\Gamma_{1}}&=\frac{Gm \rho}{r^{2}}\int_{0}^{r} \frac{K^{n,\ell}_{S_{5/3},\Gamma_{1}}}{P}dr + 4 \pi r^{2} \rho \int_{r}^{R}\frac{G\rho}{\tilde{r}^{2}}\left[\int_{0}^{\tilde{r}} \frac{K^{n,\ell}_{S_{5/3},\Gamma_{1}}}{P}d\bar{r}\right] d\tilde{r} \nonumber \\ & -\frac{5}{3}K^{n,\ell}_{S_{5/3},\Gamma_{1}}, \label{EqIntKerS}
\end{align}
and the additional boundary condition imposed on the new structural kernels is that they satisfy this equation at one point. This can be done either iteratively or by decomposing the problem into a homogeneous component and a non-homogeneous component \citep[see][for further details]{BuldgenKer}. 

\begin{figure*}[t]
	\centering
		\includegraphics[width=17.5cm]{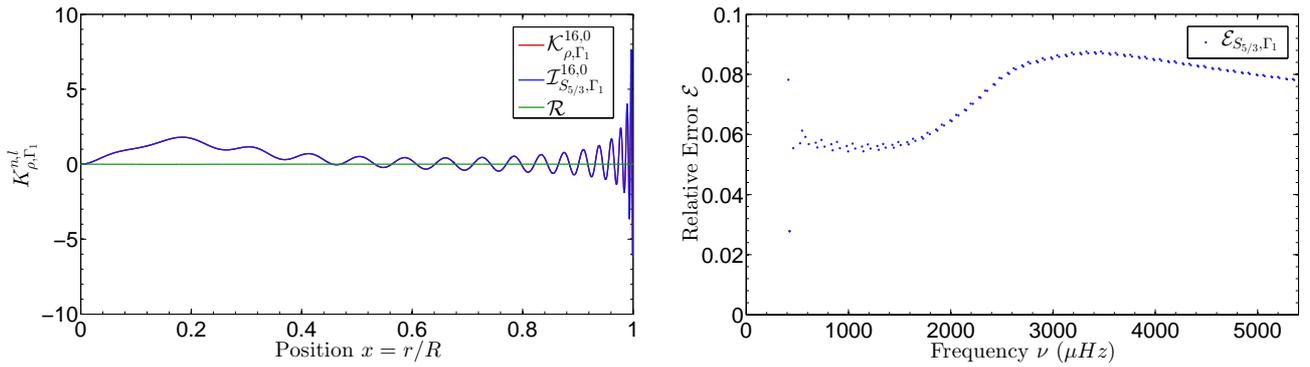}
	\caption{(Left panel) Verification of Eq. \ref{EqIntKerS} for Target $1$ of Table \ref{tabStrucTarget} for the $n=16,\ell=0$ mode kernel $K_{S_{5/3},\Gamma_{1}}^{16,0}$, where $I^{16,0}_{S_{5/3},\Gamma_{1}}$ is the right-hand side of this Equation and $\mathcal{R}$ is the residual. All quantities are presented in their non-dimensional form. (Right panel) Verification of Eq. \ref{EqVarPrinc} for modes of degree $\ell=0,1,2,3$ and various radial orders between Target $1$ of Table \ref{tabStrucTarget} and  a model of the same evolutionary sequence, $500$ $My$ younger.}
		\label{figverifSG}
\end{figure*}

\begin{figure*}[t]
	\centering
		\includegraphics[width=17.5cm]{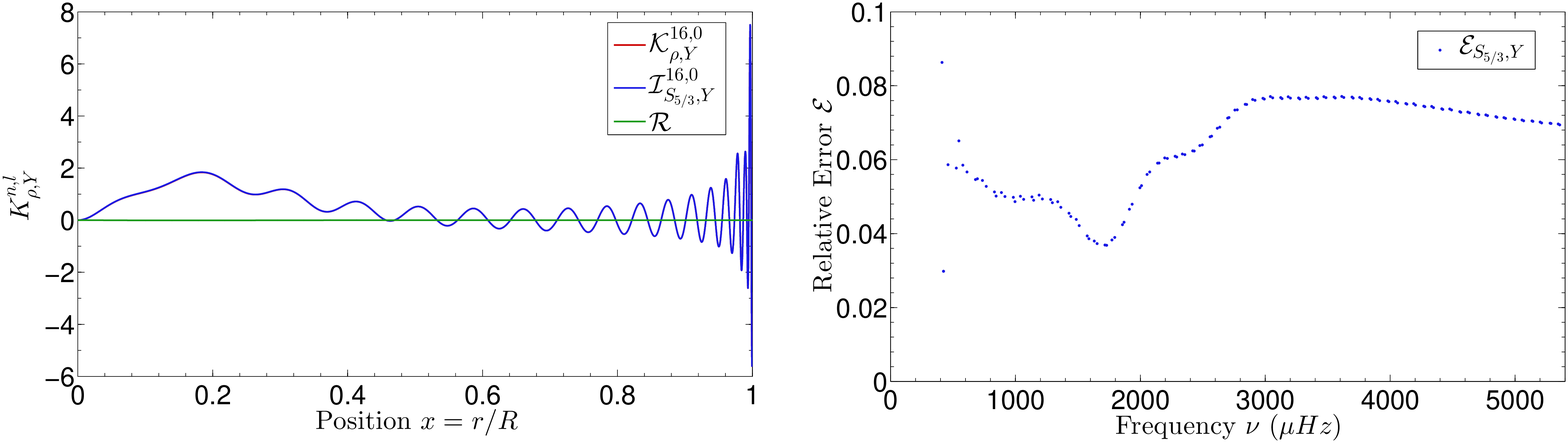}
	\caption{(Left panel) Verification of Eq. \ref{EqIntKerS} for the $(S_{5/3,Y})$ structural pair for $n=16,\ell=0$ mode kernel $K_{S_{5/3},Y}^{16,0}$, where $I^{16,0}_{S_{5/3},Y}$ is the right-hand side of this Equation and $\mathcal{R}$ is the residual. All quantities are presented in their non-dimensional form. (Right panel) Verification of Eq. \ref{EqVarPrinc} for the $(S_{5/3,Y})$ structural pair for modes of degree $\ell=0,1,2,3$ and various radial orders. The models used are the same as in Fig. \ref{figverifSG}.}
		\label{figverifSY}
\end{figure*}

Once the kernels are derived, Eq. \ref{EqIntKerS} also provides a first verification step to ensure that the kernels are in agreement with the initial steps of their derivation. The second verification is to ensure that the variational integral expressions are satisfied by the new kernels. In other words, ensure that we have:
\begin{align}
\frac{\delta \nu^{n,\ell}}{\nu^{n,\ell}}&= \int_{0}^{R}K^{n,\ell}_{\rho, \Gamma_{1}} \frac{\delta \rho}{\rho} dr + \int_{0}^{R}K^{n,\ell}_{ \Gamma_{1}, \rho} \frac{\delta \Gamma_{1}}{\Gamma_{1}} dr \nonumber \\
&= \int_{0}^{R}K^{n,\ell}_{S_{5/3}, \Gamma_{1}} \frac{\delta S_{5/3}}{S_{5/3}} dr + \int_{0}^{R}K^{n,\ell}_{ \Gamma_{1}, S_{5/3}} \frac{\delta \Gamma_{1}}{\Gamma_{1}} dr. \label{EqVarPrinc}
\end{align}
Both verifications are illustrated in Fig. \ref{figverifSG} for the $(S_{5/3},\Gamma_{1})$ pair and Fig. \ref{figverifSY} for the $(S_{5/3},Y)$ pair. The order of magnitude of the agreement is very similar to what is found for classical kernels such as the $(\rho,c^{2})$ and the $(\rho,\Gamma_{1})$ structural pairs. However, small differences can always be seen when changing the structural pair, as presented in \citet{BuldgenKer}. We recall that to ensure the verification of the variational equations, the scaling method mentioned in \citet{BuldgenKer} is of course necessary since the observed target and the reference model may not have the same radius. 

\section{Using the entropy proxy to obtain indicators of convective regions} \label{NewIndic}
In this Section, we illustrate the sensitivity of the entropy proxy, $S_{5/3}$, to convective regions and we show how to build seismic indicators useful for asteroseismic targets using the newly derived kernels. 

Inverting an integrated quantity using the SOLA inversion technique actually consists in finding the linear combination of frequency differences which best reproduces the correction to be applied to this integrated quantity. As noted before, the indicator is optimized to probe certain parts of the stellar structure. We first present the generic case of a quantity $A$, the linear perturbations of which are given by
\begin{align}
\frac{\delta A}{A} = \frac{A_{Obs}-A_{Ref}}{A_{Ref}}=\int_{0}^{1}\mathcal{T}_{A}\frac{\delta s_{1}}{s_{1}}dx, \label{EqADef}
\end{align}
with $\mathcal{T}_{A}$ being the target function used for the inversion and $s_{1}$ a variable such as $\rho$, $c^{2}$ or $S_{5/3}$. We reiterate that a generic expression of the linear integral relations can be written.
\begin{align}
\frac{\delta \nu^{n,\ell}}{\nu^{n,\ell}}&= \int_{0}^{R}K^{n,\ell}_{s_{2}, s_{1}} \frac{\delta s_{2}}{s_{2}} dr + \int_{0}^{R}K^{n,\ell}_{ s_{1}, s_{2}} \frac{\delta s_{1}}{s_{1}} dr,
\end{align}
with $s_{2}$ being the second variable of the integral relations and the $K^{n,\ell}_{s_{i},s_{j}}$ being the kernel functions related to each structural variable.

The inversion process estimates the differences between the observed integrated quantity, $A_{\mathrm{Obs}}$ and that of the reference model, $A_{\mathrm{Ref}}$, using the following relation
\begin{align}
\left(\frac{\delta A}{A}\right)_{\mathrm{Inv}}=\sum_{i}c_{i}\frac{\delta \nu_{i}}{\nu_{i}}, \label{EqAInv}
\end{align}
with $\delta \nu=\nu_{Obs}-\nu_{Ref}$ and the $c_{i}$ being the inversion coefficients determined from the minimization of the SOLA cost function \citep[see][and Eq. \ref{eqCostSola}]{Pijpers}. The inverted difference in Eq. \ref{EqAInv} is not exactly the real difference in Eq. \ref{EqADef}, due to the intrinsic limitations of the inversion.

The inversion process as a whole is a trade-off between various antagonistic terms. This is understood from the cost function used to carry out the SOLA inversion
\begin{align}
\mathcal{J}_{A} = &\int_{0}^{1}\left[ K_{\mathrm{Avg}}-\mathcal{T}_{A}\right]^{2}dx +\beta \int_{0}^{1}K^ {2}_{\mathrm{Cross}}dx + \tan(\theta) \sum^{N}_{i}(c_{i}\sigma_{i})^{2} \nonumber \\
 &+ \eta \left[ \sum^{N}_{i}c_{i}-k \right], \label{eqCostSola}
\end{align}
which contains four terms. The first term is associated with fitting the target function: it ensures the accuracy of the inversion and includes the averaging kernel, defined as
\begin{align}
K_{\mathrm{Avg}}=\sum_{i}^{N}c_{i}K^{i}_{s_{1},s_{2}},
\end{align}
with $N$ the total number of individual observed frequencies. The second integral deals with the cross-term kernel, defined as
\begin{align}
K_{\mathrm{Cross}}=\sum_{i}^{N}c_{i}K^{i}_{s_{2},s_{1}},
\end{align}
which is a pollution term by the variable $s_{2}$ whose contribution has to be damped. The third term is linked with the propagation of observational error bars of the individual frequencies, denoted here $\sigma_{i}$. The last term is linked to the additional normalization constraint on parameter $k$ derived using homologous relations (see Sect. \ref{SectNonLin} and \citep{Reese} for a description of this approach) and unlike $\beta$ and $\theta$, $\eta$ is not a free parameter of the inversion but a Lagrange multiplier. The free parameters of the inversion, $\beta$ and $\theta$, are used to give more or less importance to each of the antagonistic terms and are thus called the trade-off parameters.

The optimal set of parameters is defined by analyzing the amplitude of the first three individual terms of Eq. \ref{eqCostSola} and visually verifying that the fit is reliable. In practice, this implies that the inversion will be a compromise between precision, accuracy, and cross-term. As stated in Sect. \ref{NewKernels}, the cross-term is damped by a suitable choice of variables which has been validated in previous studies \citep{BasuAster, Reese}. Consequently, the main part of the trade-off problem is to find a suitable compromise between precision and accuracy. This is done by comparing the terms responsible for these characteristics in so-called trade-off curves \citep[see][for the full definition of this concept and various applications]{Backus,Pijpers,Rabello,Reese}, where one plots the following quantities

\begin{align}
\vert \vert K_{\mathrm{Avg}} \vert \vert^{2}=\int_{0}^{1}\left[ K_{\mathrm{Avg}}-\mathcal{T}_{A}\right]^{2}dx, \\
\sigma_{Inv}=\sqrt{\sum^{N}_{i}(c_{i}\sigma_{i})^{2}},
\end{align}
for various values of $\theta$. In practice, a visual inspection of the agreement of the averaging kernels with their target is also informative on the quality and reliability of the inverted results. In practice, the set of inversion parameters will depend on the target function of the inversion and the observed seismic data. 

In addition to these reliability assessments, further analyses can be performed, by separating the contribution to the linear correction derived from the frequencies using Eq. \ref{EqAInv}. Indeed, inversion techniques are subject to multiple error sources which can sometimes damp each other. To analyze this potential compensation, the linear correction of a given integrated quantity $A$ can be decomposed into various contributions. To do this, we introduce the quantities $A_{Ref}$, $A_{Obs}$ and $A_{Inv}$; the reference, observed and inverted values of the indicator, respectively. For these quantities, one has the following relations
\begin{align}
\frac{A_{Inv}}{A_{Ref}} &= 1 + \int_{0}^{1}K_{Avg}\frac{\delta s_{1}}{s_{1}}dx + \int_{0}^{1}K_{Cross}\frac{\delta s_{2}}{s_{2}}dx + \epsilon_{Res}, \\
\frac{A_{Obs}}{A_{Ref}} &= 1 + \int_{0}^{1}\mathcal{T}_{A}\frac{\delta s_{1}}{s_{1}}dx.
\end{align}
By computing the difference between the inverted and observed values of the indicator, one can isolate the errors in individual contributions, such that one as
\begin{align}
\frac{A_{Obs}-A_{Inv}}{A_{Ref}}=&\int_{0}^{1}\left( \mathcal{T}_{A}-K_{Avg}\right)\frac{\delta s_{1}}{s_{1}}dx - \int_{0}^{1} K_{Cross} \frac{\delta s_{2}}{s_{2}}dx- \epsilon_{Res} \nonumber \\
=&-\epsilon_{Avg}-\epsilon_{Cross}-\epsilon_{Res},
\end{align}
where we define three main error contributions: $\epsilon_{\mathrm{Avg}}$, the error stemming from the mismatch between the averaging kernel and its target, which depends on the quality of the dataset and the value of the $\theta$ parameter; $\epsilon_{\mathrm{Cross}}$, the error stemming from the non-zero cross-term contribution which depends on the choice of the variable $s_{2}$ and the parameter $\beta$; and $\epsilon_{\mathrm{Res}}$, the residual error, which is defined as
\begin{align}
\epsilon_{\mathrm{Res}}&=\frac{A_{\mathrm{Inv}}-A_{\mathrm{Obs}}}{A_{\mathrm{Ref}}}-\epsilon_{\mathrm{Avg}}-\epsilon_{\mathrm{Cross}}.
\end{align} 
This contribution is the most difficult to assess, since it can originate from surface effects, the linearization of the equation of state, or from non-linear effects. In inversions of observed targets, the residual errors would also be influenced by physical processes not included in the derivation of the variational relations and on systematics in the frequency determinations. However, since its calculation requires the knowledge of the structural differences between the target and the reference model, the value of this residual error is not accessible in practical cases. As we see in Sect. \ref{SecNumerics}, these error contributions are used in combination with the classical trade-off analysis to determine the degree of reliability of the inversion and its accuracy.

\subsection{Convective cores and deep regions}\label{IndicCore}
Convective cores represent one of the major difficulties when studying the evolution of stars with masses higher than approximately $1.2M_{\odot}$. Indeed, their presence can lead to large uncertainties in age determinations and can completely change the evolutionary track of a given model. Various studies focus on the uncertainties linked to convective cores; for example the recent studies by \citet{DeheuvelsOv} and \citet{Claret} to calibrate overshooting, or the derivation of dedicated seismic indices to the detection of a convective core in a given star \citep[See][]{MiglioAlpha}.

Looking at structural profiles of a model, the presence of a convective core can easily be seen in derivatives, where it introduces a discontinuity. This means that indicators based on derivatives, such as the $t_{u}$ indicator presented in \citet{Buldgentu}, defined as
\begin{align}
t_{u}=\int_{0}^{R}r(r-R)^{2}\exp^{-7\left(\frac{r}{R}\right)^{2}}\left(\frac{du}{dr}\right)^{2}dr,
\end{align}
 or $t$ presented in \citet{Buldgentau}, defined as 
\begin{align}
t=\int_{0}^{R}\frac{1}{r}\frac{dc}{dr}dr,
\end{align} 
will be extremely sensitive to convective cores. However, the target function associated with these indicators is impossible to fit with the structural kernels if a convective core is present in the model. Indeed, both of the target functions for $t$ and $t_{u}$ are strongly discontinuous whereas the structural kernels are not. Using $S_{5/3}$ as the structural variable for the kernels, we recover the sensitivity to convective cores without the need for a derivative. Moreover, as can be seen in Fig. \ref{figKerS}, the structural kernels associated with $S_{5/3}$ have an increased intensity in the deep regions, around $0.2 r/R$, due to the mass dependency of the entropy proxy.

We define a new indicator for convective cores as follows:
\begin{align}
S_{core}= \int_{0}^{R} \frac{f(r)}{S_{5/3}} dr, \label{EqIndicSCore}
\end{align}
with $f(r)$ being the weight function associated with the indicator: 
\begin{align}
f(r)=& r \left( a_{1} \exp \left(-a_{2}\left( \frac{r}{R}-a_{3} \right)^{2} \right) + a_{4} \exp \left(-a_{5}\left(\frac{r}{R}-a_{6} \right)^{2}\right) \right) \nonumber \\
&\tanh \left(a_{7} \left(1-\frac{r}{R} \right) \right). \label{eqRefTarCore}
\end{align}
The target function is built to fit the lobe around $0.2r/R$ present in every kernel of either the $(S,\Gamma_{1})$ or the $(S,Y)$ structural pair. It has seven non-dimensional free parameters that can be varied if required. We give an example of values for these parameters in Table \ref{TabValueParamSCore}. Their values will depend on the stellar type, because variations of the reference model of a given star will imply a slightly different behavior of the structural kernels. For example, more massive or evolved stars will tend to present a steeper decrease of the lobe around $0.2r/R$, inducing a slight increase of the parameters $a_{2}$ and $a_{5}$. The asymmetry of the lobe in the kernels can also be fitted by varying the parameters $a_{1}$ and $a_{4}$, which is also a function of the stellar type. Moreover, the parameters $a_{3}$ and $a_{6}$ might be reduced because the maximum of the lobe will be located towards deeper regions. 

\begin{table*}[t]
\caption{Example of values for the free parameters of the function $\mathcal{T}_{S_{Core}}$.}
\label{TabValueParamSCore}
  \centering
\begin{tabular}{r | c }
\hline \hline
Parameter & Example value \\ \hline 
$a_{1}$ & $10$ \\  
$a_{2}$ & $26$ \\
$a_{3}$ & $0.17$ \\
$a_{4}$ & $3$ \\
$a_{5}$ & $5$ \\
$a_{6}$ & $0.23$ \\
$a_{7}$ & $50$ \\
\hline
\end{tabular}
\end{table*}

\begin{figure}[t]
	\centering
		\includegraphics[width=9.1cm]{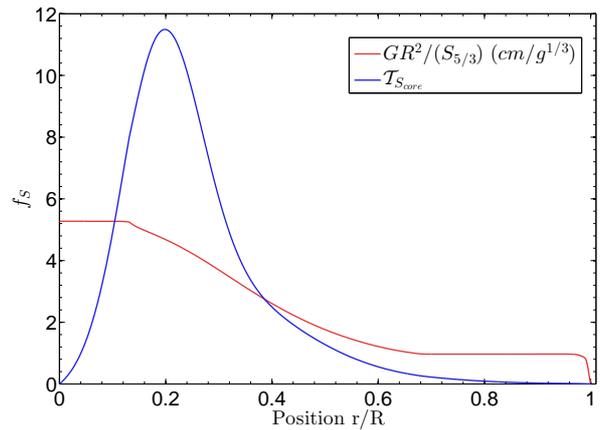}
	\caption{Target function of the core indicator $S_{Core}$ plotted alongside the structural function $S^{-1}_{5/3}$. The model used is Target $3$ of Table \ref{tabStrucTarget}. The target function is plotted in its non-dimensional form.}
		\label{figTarSCore}
\end{figure}

The linear perturbation of this indicator leads to the following target function for the inversion:
\begin{align}
\frac{\delta S_{core}}{S_{core}}= \frac{1}{S_{core}}\int_{0}^{R}-\frac{f(r)}{S_{5/3}}\frac{\delta S_{5/3}}{S_{5/3}}dr.
\end{align}
This means that the target function of this inversion is defined as
\begin{align}
\mathcal{T}_{S_{Core}}=\frac{-f(r)}{S_{Core}S_{5/3}(r)}. 
\end{align}
This function is illustrated in blue in Fig. \ref{figTarSCore} along with the profile of $1/S_{5/3}$ in red. The product of both curves gives the argument of the integral defining the indicator. From visual inspection, it is clear that this argument probes the inner layers of the acoustic structure of the star. One should also note that this indicator is not restricted to stars with convective cores but can also be a complement to or a replacement for the $t_{u}$ indicator. 

\begin{figure*}[t]
	\centering
		\includegraphics[width=17.5cm]{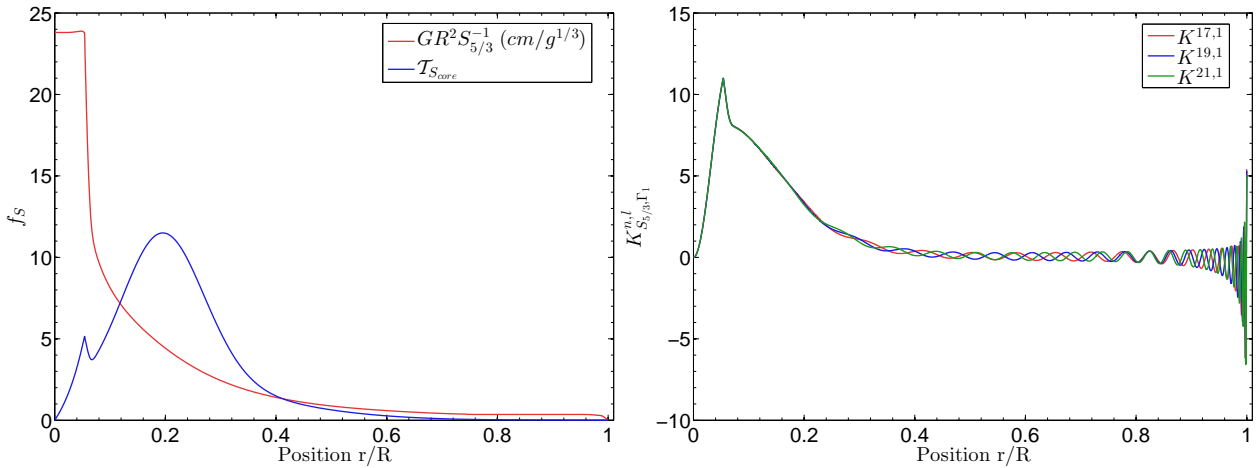}
	\caption{(Left panel) Structural profile of $S^{-1}_{5/3}$ in red, showing the plateau in the convective core. In blue, an example of the adaptation of a target function to include the trace of the convective core and its border. The target function is plotted in its non-dimensional form. (Right panel) Structural kernels associated with $S_{5/3}$ in the $(S_{5/3},\Gamma_{1})$ structural pair, showing the trace of the entropy plateau in the convective core in the central regions. The model used is Target $6$ of Table \ref{tabStrucTarget}. The kernels are plotted in their non-dimensional form.}
		\label{figTarSPeak}
\end{figure*}

If the convective core is well established and has quite a high plateau with a steep entropy variation, a peak tends to appear in the target function, as illustrated in Fig. \ref{figTarSPeak}. This peak is in fact due to the boundary of the convective core. Due to the $\mathcal{O}(r^ {2})$ behavior of the structural kernels in the center, the entropy plateau of the convective core is erased, but if its height and radial extent are sufficient, some traces remain in the kernels. Consequently the target function defined in Eq. \ref{EqIndicSCore} can still be fitted and is well adapted to extract information about convective cores. From numerical tests, the linear relations between frequencies and structure (Eq. \ref{eqKerS}) are still satisfied even in these particular cases. We see in Sect. \ref{SecNumerics} how efficient the inversion actually is when confronted with these effects. The weight functions can also be adapted to be more easily fitted if required. Ultimately, the diagnostic power is limited by the detected modes and their error bars. 
\subsection{Convective envelopes}\label{IndicEnve}

In this Section, we show that it is possible to probe the upper radiative zone and the convective envelope of certain Kepler targets. However, fully isolating the convective envelope is done thanks to high $\ell$ modes, as in helioseismology. Since these modes are not accessible to asteroseismology a complete scan of the structure is not achievable. However, a few $\ell=3$ modes can still help with extracting information from the observations. The efficiency of this technique thus depends on the ability to detect octupole modes. However, such modes are difficult to observe, even for Kepler targets, and their frequencies are often determined with a rather low precision.

\begin{figure}[t]
	\centering
		\includegraphics[width=8.5cm]{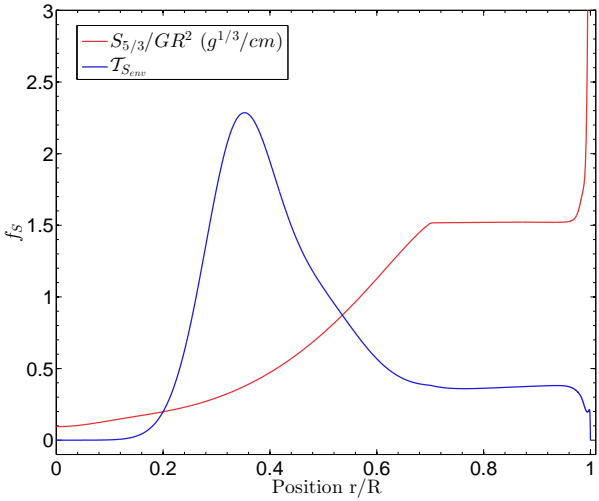}
	\caption{Target function of the envelope indicator $S_{Env}$ plotted alongside the structural function $S_{5/3}$ for Target $1$ of Table \ref{tabStrucTarget}. The target function is plotted in its non-dimensional form.}
		\label{figTarFuncEnv}
\end{figure}

Our objective here is to probe the plateau in the convective envelope of $S_{5/3}$, shown in Fig. \ref{figTarFuncEnv}, and layers immediately below it. However, a compromise has to be found to exclude the uppermost region of the star, subject to surface uncertainties and a breakdown of the assumptions behind inversion techniques. Provided that the data quality is sufficient, the following indicator can be fitted
\begin{align}
S_{env}=\int_{0}^{R}g(r)S_{5/3} dr, \label{eqSEnv}
\end{align}
which means that the target function for this inversion will be
\begin{align}
\mathcal{T}_{S_{env}}=\frac{g(r)S_{5/3}(r)}{S_{env}},
\end{align}
stemming from the relative linear perturbation of Eq. (\ref{eqSEnv}) and not forgetting the constant factor $S_{env}$ in the denominator. The weight function $g(r)$ is somewhat complicated and has been built to match the behavior of kernels while trying to extract information in the upper regions. To explain the choice of $g(r)$, we separate it into three components:
\begin{align}
g(r)=\left( g_{1}(r)+g_{2}(r) \right)g_{3}(r),
\end{align}
with the following definitions:
\begin{align}
g_{1}(r)=&r^{b_{1}}\left(b_{2}\exp \left(-b_{3}\left(\frac{r}{R}-b_{4} \right)\right)^{2} \right. \nonumber \\
& \left. +b_{5} \exp \left(-b_{6}\left(\frac{r}{R}-b_{7}\right)^{2}\right)\right), \\
g_{2}(r) =&\frac{b_{8}r^{b_{9}}}{\exp \left( \left(\frac{R}{r}-b_{10} \right)/b_{11} \right)+1}, \\
g_{3}(r)=&\tanh\left(b_{12} \left(1.0-\frac{r}{R}\right) \right). \label{eqRefTarEnv}
\end{align}
The function $g_{1}$ is used to probe deeper regions where the entropy is influenced by the way it is reconnected to the plateau of the convective envelope. Looking at Fig. \ref{figTarFuncEnv}, we see that this corresponds to the regions where a slope starts in the entropy profile. The $g_{2}$ component is a Fermi-Dirac distribution that reproduces the entropy plateau in the convective envelope. The $g_{3}$ function eliminates the surface regions as efficiently as possible through the hyperbolic tangent. This component is steep in order to avoid affecting the lower part of the profile. The target function combining all three components is represented in Fig. \ref{figTarFuncEnv}. These components can be further adjusted depending on the reference model and the observed modes. The total target function of the $S_{Env}$ indicator thus counts $12$ non-dimensional free parameters. We illustrate some examples of values for these parameters in Table \ref{TabValueParamSEnv}. All of these parameters can vary depending on the stellar type. For instance, an increase in stellar mass will reduce the size of the convective envelope and thus lead to a more difficult fit of the target function of the $S_{Env}$ indicator. This implies that the parameter $b^{8}$ to $b_{11}$ will be influenced since the plateau of $S_{5/3}$ will be located at a different place. This also implies that the $g_{1}$ function has to be adapted. The peak has to be made larger and more asymmetrical with a less steep transition towards the convective envelope position, where the $g_{2}$ sees its amplitude increase. In practice, the parameters that vary the most are the parameters $b_{2}$ to $b_{7}$. It should be noted that the peak cannot be simply shifted, because of the low degree of the modes, the maximum of the target function will always be located around $0.4r/R$ which is the highest point in normalized radius where a form of localization of the kernels can be achieved.
\begin{table*}[t]
\caption{Example of values for the free parameters of the function $\mathcal{T}_{S_{Env}}$.}
\label{TabValueParamSEnv}
  \centering
\begin{tabular}{r | c }
\hline \hline
Parameter & Example value \\ \hline 
$b_{1}$ & $1.5$ \\  
$b_{2}$ & $30$\\
$b_{3}$ & $120$ \\
$b_{4}$ & $0.3$ \\
$b_{5}$ & $7$ \\
$b_{6}$ & $45$ \\
$b_{7}$ & $0.37$ \\
$b_{8}$ & $0.4$ \\
$b_{9}$ & $1.5$ \\
$b_{10}$ & $1.7$ \\
$b_{11}$ & $1.2$ \\
$b_{12}$ & $50$ \\
\hline
\end{tabular}
\end{table*}

For example, we see in Sec. \ref{SecNumerics}, that octupole modes are required to carry out inversions of the $S_{Env}$ indicator. The number of these modes and their error bars also influence the paramaters of the $g_{1}$ function. A larger number of precisely determined octupole modes allows for a higher position of the peak in the $g_{1}$ function and a more slowly decreasing slope towards the convective envelope, implying a better analysis of these regions. Besides the observed modes, the reference model also affects the building of the target function of the $S_{Env}$ indicator. For example, a more massive star, having a shallower convective envelope, will require an adaptation of the $g_{2}$ and $g_{1}$ functions to avoid building a target that cannot be easily fitted with only low $\ell$ modes. 
\subsection{Non-linear generalization}\label{SectNonLin} 

As in \citet{Buldgentu} and \citet{Reese}, one can improve the inversion by using a simple non-linear generalization of the inversion procedure. To do this, we follow the procedure in Sect. $3.2$ of \citet{Buldgentu}. From homology scaling, we can show that
\begin{align}
S_{5/3} = \frac{P}{\rho^{5/3}}\propto \frac{GM^{2}R^{5}}{M^{5/3}R^{4}}\propto GRM^{1/3} \label{DimEq}.
\end{align}
This homology scaling is also crucial to know how the indicator value should be rescaled to be comparable for various reference models. We have seen in \citet{Buldgentu} that the inversion implicitly rescales the target model to the radius of the reference model without changing its mean density. This means that the quantity $S_{5/3}$ is then also rescaled. Consequently, when comparing inversions from various reference models, one should always be aware of this scaling and present them in a form proportional to the mean density. For example, using Eqs. \ref{EqIndicSCore} and \ref{eqRefTarCore}, one can carry out a homology scaling of the core conditions indicator, leading to
\begin{align} 
S_{Core}\propto \frac{R^{2}}{S_{5/3}} \propto \frac{R}{GM^{1/3}},
\end{align}
meaning that the quantity $GS_{Core}$ scales as $\rho^{-1/3}$. For $S_{Env}$, homology scaling from Eqs. \ref{eqSEnv} and \ref{eqRefTarEnv} leads to
\begin{align}
S_{Env}\propto R^{3.5}M^{1/3}G.
\end{align}
Meaning that this quantity needs to be rescaled if it is to be compared for various reference models. In other words, one has $S_{Env}/GR^{4.5}\propto \rho^{1/3}$. To provide the coefficients for the non-linear generalization, we focus on the mass dependency of the quantities, since the radius is implicitly kept constant by the inversion. Looking at Eq. \ref{DimEq}, this means that $S_{5/3} \propto \nu^{2/3}$ in terms of the $M$ dependency, thus leading to
\begin{align}
\frac{\delta S_{5/3}}{S_{5/3}}=\frac{2}{3}\frac{\delta \nu}{\nu},
\end{align}
for a homologous transformation which keeps the radius between models constant but changes their mean density through their mass. As stated in Sect. $3.2$ of \citet{Buldgentu}, the coefficient in this linear relation between perturbations of the frequencies and that of a structural quantity, denoted $k$ in our previous paper, is crucial to derive the non-linear generalization. In this study, one has $k=2/3$ in the relation between $S_{5/3}$ and the frequencies. Looking at Eq. \ref{eqSEnv}, which defines $S_{env}$, we can see that since this indicator is proportional to $S_{5/3}$. Hence, one will have

\begin{align}
\frac{\delta S_{Env}}{S_{Env}}=\frac{2}{3}\frac{\delta \nu}{\nu},
\end{align}

and the value of $2/3$ can be applied to Eq. $31$ and $34$ of \citet{Buldgentu} defining the optimal value of $S_{Env}$ and its associated errors bars in the framework of this non-linear generalization. Similarly, the value $k$ will also be fixed to $2/3$ in the additional condition on the inversion coefficients (fourth term of Eq. \ref{eqCostSola}) used to improve the regularization of the inversion process.

Using the same analysis, it is easy to show that the $S_{Core}$ indicator will satisfy a relation of the opposite sign, due to its opposite mass dependency,
\begin{align}
\frac{\delta S_{Core}}{S_{Core}}=\frac{-2}{3}\frac{\delta \nu}{\nu},
\end{align}
since it is proportional to $S^{-1}_{5/3}$. Consequently, the value $-2/3$ can also be used in Eqs. $31$, $34$ and $28$ of \citet{Buldgentu} to derive the non-linear generalization of this indicator.

\subsection{Relation between $S_{core}$, $S_{env}$, and stellar structure}
In the preceding Sections, we have shown how the entropy proxy could be used to obtain indicators of both convective cores and envelopes. In this Section we will briefly show how they change with some specific aspects of stellar structure. We mention that these changes of course depend on the parameters used to build the target functions of the indicators. For example, placing the maximum of the Gaussian functions deeper in the $S_{Core}$ target function will increase the changes due to convective cores. 

\begin{figure*}[t]
	\centering
		\includegraphics[width=15.5cm]{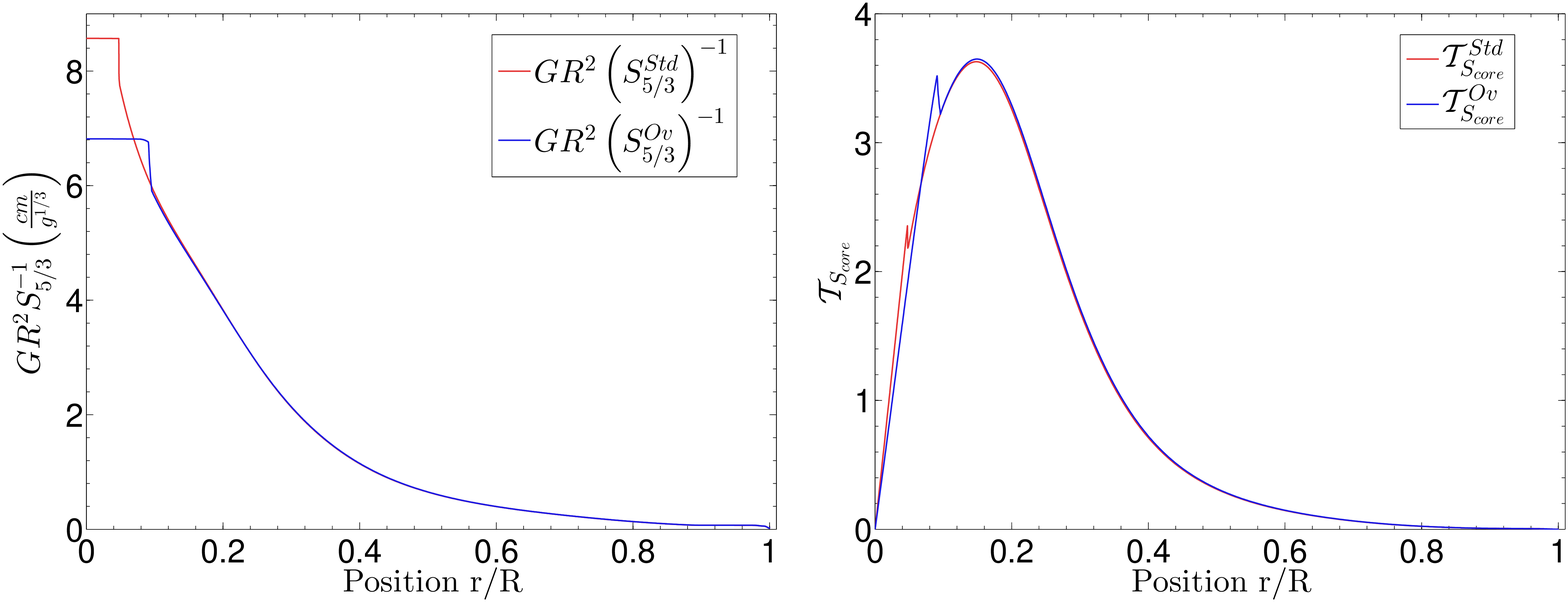}
	\caption{(Left panel) Scaled $S_{5/3}^{-1}$ profile used in inversions for two $1.3M_{\odot}$ models. The blue curve is associated with a model built using a $0.15$ pressure scale height adiabatic overshoot. (Right panel) Target function of the $S_{core}$ indicator for both models.}
		\label{figSCoreStruc}
\end{figure*}

We illustrate the relation between convective cores and the $S_{core}$ indicator in Fig. \ref{figSCoreStruc} by plotting the rescaled $S^{-1}_{5/3}$ profiles of two $1.3M_{\odot}$, $1.6Gy$ models with a solar chemical composition. The model plotted in blue in Fig. \ref{figSCoreStruc}, included a $0.15$ pressure scale-height adiabatic overshoot, causing its convective core to be larger at a given age and inducing changes in its entropy plateau. Similarly, the peak in the $S_{core}$ target function is much more pronounced in the blue profile than in the red, indicating a more extended convective core. The changes seen in the indicator thanks to this small variation are of around one percent of the total value of the indicator. This is of course quite small but results from the fact that both models have the same age and chemical composition, and only differ in one physical ingredient. In a more realistic case, where the models are selected using seismic constraints, the differences between reference model and target can be much larger.

\begin{figure*}[t]
	\centering
		\includegraphics[width=15.5cm]{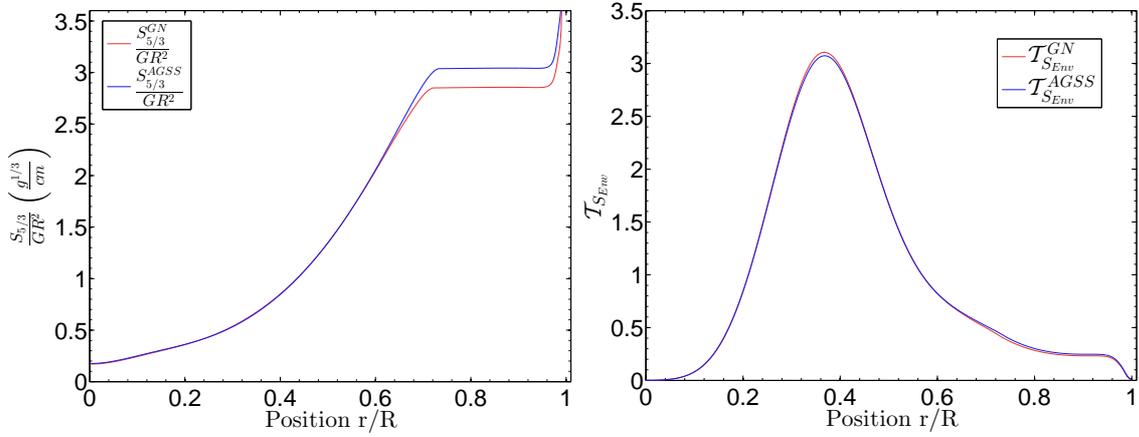}
	\caption{(Left panel) Scaled $S_{5/3}$ profile used in inversions for two standard solar models. The blue curve is associated with a model built using the AGSS09 abundances while the red one uses the GN93 abundances. (Right panel) Target function of the $S_{env}$ indicator for both models.}
		\label{figSEnvStruc}
\end{figure*}

Similarly, we illustrate in Fig. \ref{figSEnvStruc} the effects of opacity changes on solar models. We compare solar models using the AGSS09 and the GN93 abundances and different opacity tables. To further increase the differences, we do not include microscopic diffusion in the AGSS09 solar model. This leads to variations in the position of the entropy plateau of the convective envelope of the models, as can be seen in the left panel of Fig. \ref{figSEnvStruc}. This emphasizes the direct link between the entropy plateau and the temperature gradient in stellar models. The right panel of this Figure illustrates the small changes in the target function of $S_{Env}$. The changes in the target function induce a change in the value of the indicator of less than one percent. Again this is quite small but is mainly due to the fact that the models have the same mass, the same age and are not selected on the basis of their seismic information. Sect. \ref{SecNumerics} shows a very different picture for its first target, a solar model including various physical processes fitted using individual small frequency separations without including these additional processes in the reference model.

\section{Tests on artificial data}\label{SecNumerics}
In this Section, we present the results of exercises carried out to test the accuracy and the capabilities of the indicators based on the entropy proxy. We test these capabilities in similar exercises as in \citet{Buldgentu}, using targets built with certain specificities that the reference models of the inversion did not include. For example, some targets included strong overshooting or turbulent diffusion and were fitted with reference models without these processes. This approach attempts to simulate as best as possible the biases due to physical simplifications or approximations in our representation of stellar structure and evolution when carrying out seismic modeling of observed targets. It is also a good test of robustness and accuracy to see whether the linear approximation and the SOLA method are indeed capable of probing such aspects of stellar modeling.

In total, we built ten targets with various masses and ages, but only present six to avoid redundancy. The reference models were built by fitting the targets' individual small frequency separations and their effective temperatures. The fit was carried out using a Levenberg-Marquardt algorithm for the minimization. The observational spectrum of the targets is given in Table \ref{tabFreqTarget}. An error bar of $80K$ was used on the effective temperature and the observational error bars on the frequencies were similar to those found for the best Kepler targets. The physical characteristics of the target models are given in Table \ref{tabStrucTarget}. Some parameters have rather extreme values, in order to test the inversion and its limitations. Targets $1$ to $5$ were built using the OPAS opacities \citep{Mondet,Lepennec} and the Ceff equation of state \citep{CEFF} with the AGSS09 abundances \citep{AGSS} while Target $6$ was built using the OPAL opacities \citep{OPAL} and the OPAL equation of state \citep{OPALEOS} with its corrections \citep{Rogerseos}. The reference models all used the Ceff equation of state and the OPAL opacities along with the GN93 abundances \citep{GN93}. None of the reference models included turbulent diffusion. The formalism used to implement turbulent diffusion in the target models is presented in \citet{Miglio}. All models were computed using the Liège Stellar Evolution Code \citep[Cles,]{ScuflaireCles} and the oscillations were computed using the Liège Oscillation Code \citep[Losc,]{ScuflaireLosc}. The reference models obtained from this first step of forward seismic modeling are presented in Table \ref{tabHounds}. As can be seen, most of the reference models, denoted ``$Ref$'', still present significant structural differences from their respective target from Table \ref{tabStrucTarget}. The goal of the exercises is now to determine whether the inversions of the structural indicators can detect these mismatches, and thus provide additional information which could lead to refinements of the seismic modeling in the study of an observed target\footnote{Similar exercises can be found in Sect. $5$ of \citet{Buldgentu}}.

\begin{table*}[t]
\caption{Characteristics of the target models used in this study.}
\label{tabStrucTarget}
  \centering
\begin{tabular}{r | c | c | c | c | c | c}
\hline \hline
& Target$_{1}$ & Target$_{2}$ & Target$_{3}$ & Target$_{4}$ & Target$_{5}$ & Target$_{6}$\\ \hline
Mass $(M_{\odot})$ & $1.0$ & $1.1$ & $0.86$ & $1.14$ & $1.05$ &$1.07$\\
Age $(Gy)$ & $4.5$ & $6.0$ & $2.0$ & $3.5$ & $1.5$ & $5.0$\\  
Radius $(R_{\odot})$ & $0.9896$&  $1.3492$ & $0.8178$& $1.0908 $& $1.1033$ &$1.56$\\  
$T_{eff}$ $(K)$ & $5823$& $5907$ & $5056$& $5748$ & $5918$ &$5.983$\\
$Z_{0}$ & $0.02$ & $0.02$ & $0.015$ & $0.03$ & $0.02$&$0.013$\\
$X_{0}$ & $0.69$ & $0.69$ & $0.67$ & $0.7$ &$0.65$& $0.68$\\
Abundances & AGSS$09$ & AGSS$09$ &AGSS$09$ & AGSS$09$ & AGSS$09$& GN$93$\\
$\alpha_{MLT}$ & $2.0$ & $1.8$ & $1.7$ & $1.9$ & $1.5$ & $1.7$\\
$\alpha_{Ov}$ & $0.1$ & $0.2$ & $0.2$ & $0.2$ & $0.2$ & $0.15$\\
Microscopic Diffusion & Yes & Yes & Yes & Yes & Yes & Yes\\
Turbulent Diffusion & Yes & Yes & No & Yes & Yes & Yes\\
\hline
\end{tabular}
\end{table*}

\begin{table*}[t]
\caption{Characteristics of the reference models obtained from the forward modeling process.}
\label{tabHounds}
  \centering
\begin{tabular}{r | c | c | c | c | c | c}
\hline \hline
& Ref$_{1}$ & Ref$_{2}$ & Ref$_{3}$ & Ref$_{4}$ & Ref$_{5}$ & Ref$_{6}$\\ \hline
Mass $(M_{\odot})$ & $1.17$ & $1.29$ & $0.84$ & $0.93$ & $1.04$ &$1.12$\\
Age $(Gy)$ & $3.3$ & $4.6$ & $1.8$ & $4.0$ & $1.3$ & $4.5$\\  
Radius $(R_{\odot})$ & $1.05$&  $1.42$ & $0.80$& $1.02 $& $1.10$ &$1.55$\\  
$T_{eff}$ $(K)$ & $5826$& $5991$ & $5069$& $5751$ & $5960$ &$6035$\\
$Z_{0}$ & $0.052$ & $0.036$ & $0.015$ & $0.013$ & $0.014$&$0.015$\\
$X_{0}$ & $0.65$ & $0.68$ & $0.7$ & $0.7$ &$0.68$& $0.67$\\
Abundances & GN$93$ & GN$93$ &GN$93$ & GN$93$ & GN$93$& GN$93$\\
$\alpha_{MLT}$ & $2.5$ & $2.2$ & $1.4$ & $1.4$ & $1.4$ & $1.8$\\
$\alpha_{Ov}$ & $0.0$ & $0.1$ & $0.0$ & $0.1$ & $0.2$ & $0.1$\\
Microscopic Diffusion & Yes & Yes & Yes & Yes & Yes & Yes\\
Turbulent Diffusion & No & No & No & No & No & No\\
\hline
\end{tabular}
\end{table*}

We also tested the impact of supplementary observed modes on the accuracy of the inversions; for example $\ell=4$ and $5$ modes but also modes with higher and lower $n$. Our goal was to assess which oscillations modes could help with extracting more information on stellar structure using inversions of integrated quantities. We discuss this at the end of Sect. \ref{SecEnv}. The error bars were usually taken to be similar to those of the best Kepler targets, that is around $3 \times 10^{-1}$ $\mu Hz$ with slightly larger error bars for the lowest and highest frequencies, as expected in observed cases.
\begin{table*}[t]
\caption{Frequencies used to fit the simulated targets.}
\label{tabFreqTarget}
  \centering
\begin{tabular}{r | c | c | c | c}
\hline \hline
 $\ell $& $0$ & $1$ & $2$ & $3$ \\ \hline
 $n$ &$13-27$ & $13-27$ & $12-26$& $17-23$\\
\hline
\end{tabular}
\end{table*}

\subsection{Results for $S_{Core}$ inversions}
\begin{figure*}[t]
	\centering
		\includegraphics[width=17.5cm]{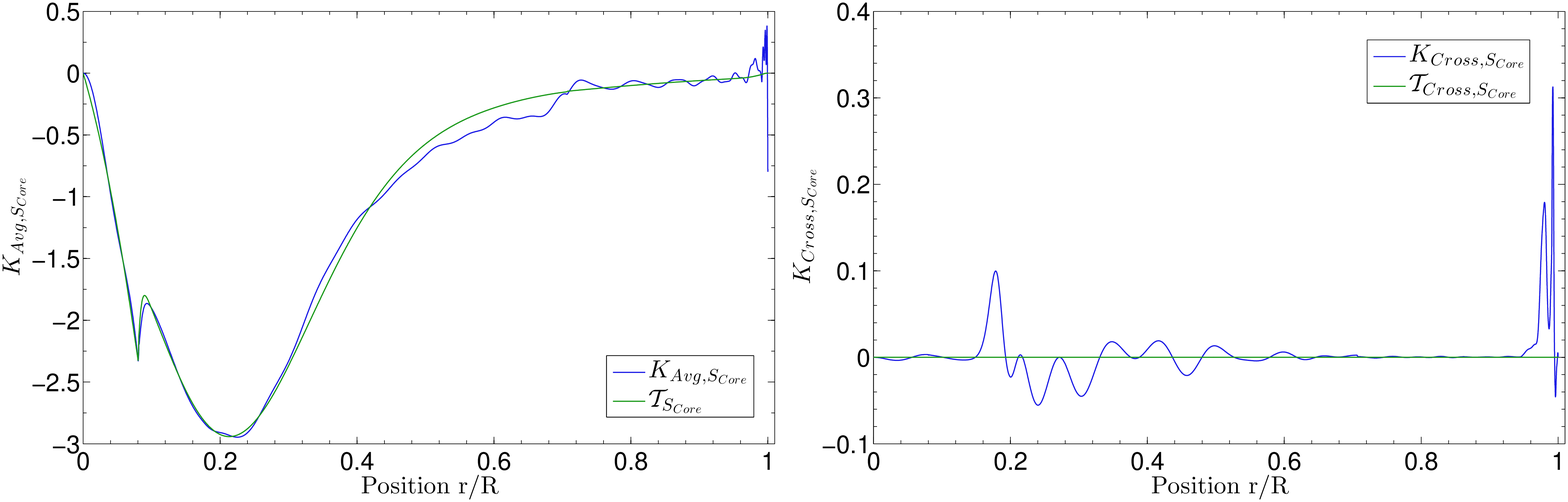}
	\caption{(Left panel) Averaging kernel for the $S_{Core}$ indicator (blue) and the target function for the inversion (green). (Right panel) Cross-term kernel for the $S_{Core}$ inversion (blue), and target function which is $0$ (green). These results are presented for Target $2$ of Table \ref{tabStrucTarget}.}
		\label{figKerSCore}
\end{figure*}

In Table \ref{tabresultsSCore}, we summarize the results of our test cases. Some kernels are illustrated in Fig. \ref{figKerSCore}. In these tests, we consider observed quantities, denoted with the subscript ``$Tar$'', to be those of the target models.

\begin{table*}[t]
\caption{$S_{Core}$ inversion results for the six targets using the $(S_{5/3}, Y)$ kernels.}
\label{tabresultsSCore}
  \centering
\begin{tabular}{r | c | c | c | c | c | c }
\hline \hline
 & $S_{Core}^{\mathrm{Ref}}G$ $(cm/g^{1/3})$ & $S_{Core}^{\mathrm{Inv}}G$ $(cm/g^{1/3})$ & $S_{Core}^{\mathrm{Tar}}G$ $(cm/g^{1/3})$ & $\varepsilon_{\mathrm{Avg}}^{S_{5/3},Y}$& $\varepsilon_{\mathrm{Cross}}^{S_{5/3},Y}$&$\varepsilon_{\mathrm{Res}}^{S_{5/3},Y}$\\ \hline
 $\mathrm{Target}_{1}$& $3.201$ & $3.118\pm 0.0251$ & $3.126$ & $-3.085\times 10^{-3}$& $-1.744\times 10^{-4}$& $7.104\times 10^{-4}$\\
 $\mathrm{Target}_{2}$ & $3.817$ &$3.660\pm 0.067$&$3.668$& $-2.477 \times 10^{-3}$& $-3.235 \times 10^{-5}$ & $1.110 \times 10^{-4}$\\
 $\mathrm{Target}_{3}$ &$2.789$ &$2.786\pm 0.034$&$2.783$& $5.714 \times 10^{-4}$& $1.868 \times 10^{-4}$& $1.783 \times 10^{-4}$ \\
  $\mathrm{Target}_{4}$ & $3.124$&$3.190\pm 0.048$&$3.184$&$1.706 \times 10^{-3}$& $-1.383 \times 10^{-4}$&$4.121 \times 10^{-4}$ \\
 $\mathrm{Target}_{5}$ & $2.734$ &$2.720\pm 0.007$&$2.726$& $-1.758 \times 10^{-3}$& $2.424\times 10^{-4}$& $-5.751\times 10^{-4}$\\
  $\mathrm{Target}_{6}$ & $3.408$ &$3.434\pm 0.003$&$3.438$& $-1.357 \times 10^{-3}$& $-5.035\times 10^{-5}$& $3.779\times 10^{-4}$\\
\hline
\end{tabular}
\end{table*}

From the inversions for Targets $2$ and $5$, we can see that the $S_{Core}$ indicator efficiently detects an inaccuracy in the stratification of the reference models, due to the absence of convective cores in these models. For Target $3$, we can see that the inversion provides an accurate result, but that the error bars on the inverted result are too large to conclude that the model has to be rejected. From a modeling point of view, we can see from Tables \ref{tabHounds} and \ref{tabStrucTarget} that Ref$_{3}$ is already a very good fit to Target $3$. Adding more frequencies to this test case allowed to eliminate this problem but in practice, this would mean that for this particular case, the $S_{Core}$ indicator would only be an additional check for the modeling. However, in most test cases, the observed differences between reference and inverted result are smaller than the error bars of the inverted result, meaning that we are still safe in terms of trade-off parameters.

Moreover, it should be noted that the values of the parameters in the definition of these indicators are not fixed. The number of frequencies available and the characteristics of the convective core (if present) imply that the parameters considered optimal in one case might be suboptimal in another case, depending on the mass, evolutionary stage or chemical composition of the model. For each of the targets in Table \ref{tabresultsSCore}, we fine-tuned the parameters to analyze the diagnostic potential using the oscillation modes of Table \ref{tabFreqTarget}. Further tests on targets of the Kepler Legacy sample \citep[See][]{ProcLeg} have also shown similar behaviors and proved that the method was indeed applicable to current seismic data.

From Table \ref{tabresultsSCore}, we can conclude that the inversion is efficient and can provide a diagnostic of inaccuracies in our modeling of deep regions of solar-like stars with convective cores. Very slight compensation is seen for Targets $5$ and $6$, but further tests have shown these cases to be marginal. In addition, the averaging kernels illustrated in Fig \ref{figKerSCore} fit their target function to an acceptable level of accuracy. 

\subsection{Results for $S_{Env}$ inversions}\label{SecEnv}

\begin{figure*}[t]
	\centering
		\includegraphics[width=17.5cm]{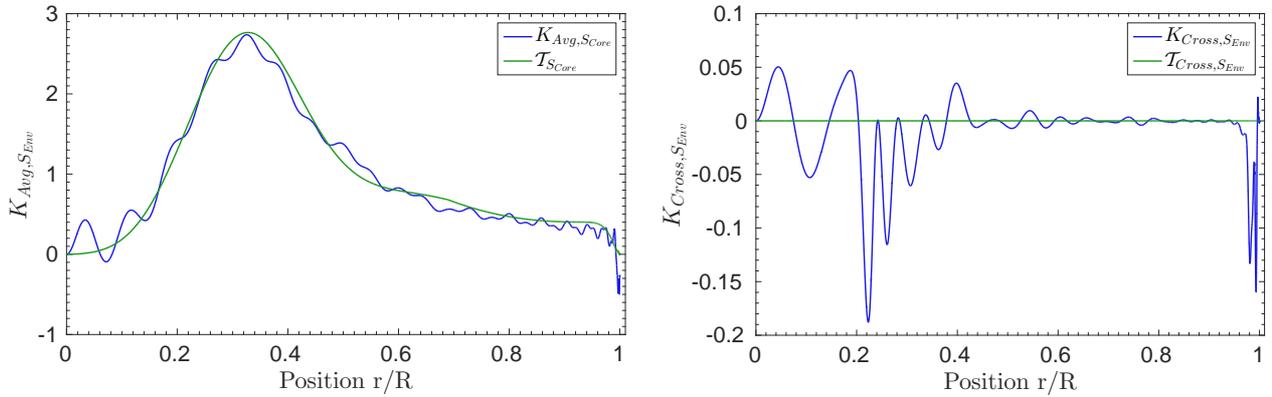}
	\caption{(Left panel) Averaging kernel for the $S_{Env}$ indicator (blue) and the target function for the inversion (green). (Right panel) Cross-term kernel for the $S_{Env}$ inversion (blue), and target function which is $0$ (green). These results are presented for Target $1$ of Table \ref{tabStrucTarget}.}
		\label{figKerSEnv} 
\end{figure*}

In Table \ref{tabresultsSEnv}, we present inversion results for the $S_{Env}$ indicator. Some kernels are shown in Fig \ref{figKerSEnv}. We used the same targets, oscillation spectra and naming conventions as for the $S_{Core}$ indicator. 

\begin{table*}[t]
\caption{$S_{Env}$ inversion results for the six targets using the $(S_{5/3}, Y)$ kernels.}
\label{tabresultsSEnv}
  \centering
\begin{tabular}{r | c | c | c | c | c | c }
\hline \hline
 & $\frac{S_{Env}^{\mathrm{Ref}}}{R^{4.5}_{Ref}G^{1/3}}$ $(g^{1/3}/cm)$ & $\frac{S_{Env}^{\mathrm{Inv}}}{GR^{4.5}_{Ref}}$ $(g^{1/3}/cm)$ & $\frac{S_{Env}^{\mathrm{Tar}}}{GR^{4.5}_{Tar}}$ $(g^{1/3}/cm)$ & $\varepsilon_{\mathrm{Avg}}^{S_{5/3},Y}$& $\varepsilon_{\mathrm{Cross}}^{S_{5/3},Y}$&$\varepsilon_{\mathrm{Res}}^{S_{5/3},Y}$\\ \hline
 $\mathrm{Target}_{1}$& $1.682$ & $1.805\pm 0.035$ & $1.807$ & $-2.079\times 10^{-3}$& $1.560\times 10^{-4}$& $4.110\times 10^{-2}$\\
 $\mathrm{Target}_{2}$ & $1.611$ &$1.759\pm 0.071$&$1.772$& $9.473 \times 10^{-4}$& $9.717 \times 10^{-6}$ & $-8.521 \times 10^{-3}$\\
 $\mathrm{Target}_{3}$ &$1.127$ &$1.050\pm 0.046$&$1.040$& $1.338\times 10^{-2}$& $2.127 \times 10^{-5}$& $-3.964 \times 10^{-3}$ \\
  $\mathrm{Target}_{4}$ & $0.924$&$0.914\pm 0.057$&$0.9083$& $5.787 \times 10^{-3}$& $-8.729\times 10^{-5}$& $-8.189 \times 10^{-4}$ \\
 $\mathrm{Target}_{5}$ & $1.626$ &$1.627\pm 0.081$&$1.629$& $-5.462 \times 10^{-4}$& $-2.837\times 10^{-5}$& $-6.224\times 10^{-4}$\\
  $\mathrm{Target}_{6}$ & $1.777$ &$1.849\pm 0.044$&$1.848$& $8.039 \times 10^{-3}$& $-1.374\times 10^{-4}$& $-7.517\times 10^{-3}$\\
\hline
\end{tabular}
\end{table*}

\begin{figure*}[t]
	\centering
		\includegraphics[width=17.5cm]{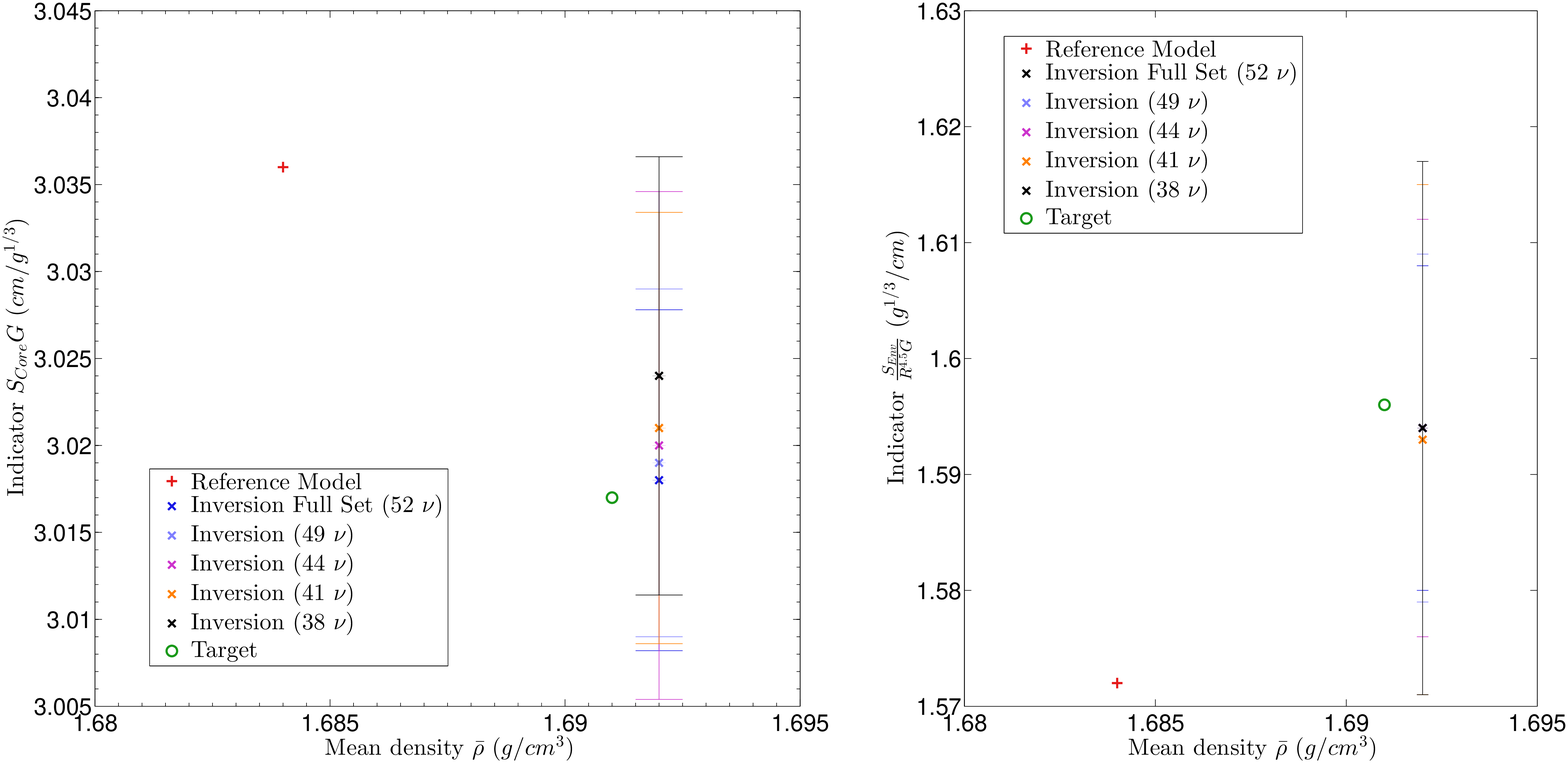}
	\caption{(Left panel) Illustration of the accuracy of the inversion of the $S_{Core}$ indicator illustrated in a $\bar{\rho}-S_{Core}$ plane for various numbers of observed frequencies. The reference values are given in red, the target values are plotted in green and the inverted values are plotted in various colors according to the number of observed frequencies. (Right panel) Same for the $S_{Env}$ indicator.}
		\label{figInvSuppl}
\end{figure*}

From Table \ref{tabresultsSEnv}, we can see that reproducing the value of $S_{Env}$ can be done efficiently for most targets. However, for Target $5$, for example, the inversion is very difficult because of the proximity between the target and its reference model. The variation of the indicator is too small to be seen with the typical accuracy of asteroseismic data. In comparison, some standard solar models show larger differences from one to another than the difference seen between Target $5$ and its reference model. In addition, Target $5$ is more massive, meaning that its envelope does not go as deep as in a lower-mass star. In fact, we found the base of the envelope to be around $0.75$ fractional radii in its reference model. Therefore, the sensitivity of the indicator has to be increased by pushing the weight function towards upper regions. However, the structural kernels, illustrated in the right panel of Fig. \ref{figTarSPeak} for a massive star with a convective core (noticeable by the peak in the deeper layers) show the exact opposite trend, the more massive the model, the steeper the decrease towards upper regions. Consequently, the amount of seismic information required to probe the convective envelope of massive stars is higher than for low-mass stars, for which the entropy plateau goes deeper and the kernels are more suited to their purpose.
We confirm this by carrying out supplementary test cases for additional low-mass targets and show that these can be probed with the $S_{Env}$ indicator. We illustrate one of these tests in Fig. \ref{figInvSuppl} and also test the effects of reducing the number of observed frequencies. For the $S_{Core}$ indicator, reducing to $44$ frequencies still works. However, by changing the parameters in the target function of the indicator, valuable information can still be gained with as low as $35$ observed frequencies. For the $S_{Env}$ indicator, using fewer than $45$ frequencies may already lead to imprecise results and we noticed that $\ell=3$ modes were required to ensure an acceptable fit of the target function. Its range of application is therefore limited to the very best observed Kepler targets. In this particular case, we used a $0.9M_{\odot}$ model including microscopic diffusion and fitted the individual large and small frequency separations of this artificial target with models that did not include microscopic diffusion and had a different helium mass fraction than the target.

Including higher $\ell$ modes can help improve the fit of the target function, but not all radial orders are equivalent in this matter. As already observed for the $t_{u}$ inversion, low $n$ modes are better at fitting custom-made global quantities, while high $n$ modes are barely used and associated with lower inversion coefficients. Indeed, at very high values of $n$, one reaches the asymptotic regime and the oscillation spectrum becomes very regular, thus the information to be extracted from the modes is degenerate. Moreover, the large error bars on these high frequencies might make them very difficult to use for precise determinations of seismic indicators and these modes are more affected by the surface effects, biasing their seismic diagnostic.
\section{Conclusion}

From Sect. \ref{SecNumerics}, we can see that both indicators are well suited to probing convective cores and envelopes. We can see from Table \ref{tabresultsSCore} that the residual error remains small compared to the variations of the $S_{Core}$ indicator caused by the changes of the entropy plateau in the convective core.  Consequently, we can conclude that the $S_{Core}$ indicator is suitable for analyzing mismatches in the deep layers of solar-like stars, even if they present a convective core. The $S_{Env}$ indicator, on the other hand, is quite efficient in its analysis of upper layers and can provide further constraints even if other indicators, such as the $t_{u}$ indicator, are fitted within their error bars.

Another aspect of inversions that has to be re-analyzed is linked to the surface effects. While inversions related to core-sensitive structural aspects are naturally less prone to showing problems, this might not be the case for all inversions. Actually, additional tests have shown that some inaccuracies, which appear as an increase of the residual error of the inversion, might be expected. Therefore, one needs to analyze how empirical corrections such as those of \citet{Sonoi} and \citet{Ball} might help solve the problem, since the classical surface-correcting method used in helioseismology cannot be used in asteroseismology due to the limited number of observed frequencies, this analysis will be carried out in future studies.

In conclusion, we have proved that the amount of seismic information found in the typical spectra of solar-like Kepler targets is sufficient to carry out more robust inversions of a core condition indicator, applicable to more massive stars. We have also shown that for the very best of these targets, for which octupole modes are observed, the inversion of a structural indicator probing the regions near the base of the convective envelope could be attempted, providing supplementary insight into the structural properties of the target. However, both $S_{Core}$ and $S_{Env}$ might require a slight adaptation of their target functions when applied to observations depending on the dataset and the reference model for the inversion. Nevertheless, they can be used alongside additional interferometric, spectroscopic or even seismic constraints such as, for example, glitch-fitting techniques. In turn, the use of these indicators will help improve the physical accuracy of stellar models and provide stringent constraints on stellar fundamental parameters required by other fields such as exoplanetology and Galactic archeology. 
\begin{acknowledgements}
G.B. is supported by the FNRS (``Fonds National de la Recherche Scientifique") through a FRIA (``Fonds pour la Formation à la Recherche dans l'Industrie et l'Agriculture") doctoral fellowship. This article made use of an adapted version of InversionKit, software developed in the context of the HELAS and SPACEINN networks, funded by the European Commission's Sixth and Seventh Framework Programmes. We thank the referee for his useful comments which have helped improve the manuscript.
\end{acknowledgements}
\bibliography{biblioarticle6}
\end{document}